\documentclass[review]{elsarticle}

\journal{Nuclear Instruments and Methods in Physics Research A}

\usepackage{amssymb}

\usepackage{amsthm}
\usepackage{physics}
\usepackage{todonotes}
\usepackage{ulem}
\usepackage{xspace}

\usepackage{lineno}

\usepackage{cleveref}
\usepackage{url}

\begin{document}

\begin{frontmatter}

\title{Optimization of the light intensity for Photodetector calibration}

\author[jinraddress]{N. ~Anfimov\corref{cor1}}
\ead{anphimov@gmail.com}
\author[jinraddress]{A. Rybnikov}
\author[jinraddress]{A. Sotnikov}
\cortext[cor1]{Corresponding author. tel. +7 496 2164126}
\address[jinraddress]{Joint Institute for Nuclear Research, Joliot-Curie 6, Dubna, Russia, 141980}

\begin{abstract}
In this article we present an evaluation of the uncertainty in the average number of photoelectrons, which is important for the calibration of photodetectors. We show that the statistical uncertainty depends on light intensity, and on the method of evaluation. For some cases there is optimal light intensity where the accuracy reaches its optimal value with fixed statistics. A method of photoelectron evaluation based on the extraction of pedestal's (zero) probability gives the best accuracy at approximately 1.6 photoelectrons for a noiseless photodetector and shifts out to higher values with the presence of noise. In the general case, estimation of the average number of photoelectrons is biased and might need special consideration.
\end{abstract}


\end{frontmatter}

\section{Introduction}
\label{sec:introduction}
Modern and future large High Energy Physics experiments exploit thousands or even dozens of thousands of photodetectors as vacuum Photo-Multiplier Tubes (PMTs)~\cite{JUNO, IceCube, BaikalGVD} or novel Silicon PhotoMultipliers (SiPM)~\cite{nEXO, PANDA, MEG-II}.
Commissioning of large batches of photodetectors requires time-efficient and robust methods of determining of their characteristics.
Photon Detection Efficiency (denoted hereafter as $\varepsilon$) is one of the key parameters of a photodetector.

There are several methods used elsewhere to estimate $\varepsilon$. 
Most of them rely on the assumption of a known mean number of photons $\overline{n}_\gamma$ hitting the photodetector.
Therefore, one attempts to determine the mean number $\mu=\overline{n}_\gamma \varepsilon$ of photoelectrons.
The widely used methods to determine $\mu$ include (i) fitting the observed charge spectrum within a photodetector's response model, (ii) estimation of the dispersion of charge distribution, (iii) estimating the occurrence probability of events with no photoelectrons ($n=0$), known as the pedestal~\cite{smirnov}, or similarly examining the events with a particular number $n\ne 0$ of photoelectrons.

Our main motivation for this work is to demonstrate that the efficiency of the method (iii) can be greatly improved for particular values of $\mu$.
For example, for the pedestal method $\mu \approx 1.6$ is found to be optimal, i.e. it requires the shortest acquisition time to reach the desired relative precision (standard deviation) $\sigma_\mu/\mu$.
This observation might be of practical usefulness when large tests of photodetectors are considered.
In particular, this method is applicable to the PMT mass test procedure described in~\cite{JUNO-anfimov}.

The paper is organized as follows. 
In~\cref{sec:statistics} we provide a short introduction to methods of statistical analysis used in this work.
Determinations of $\mu$ and $\sigma_\mu$ of a photodetector without and with noise are summarized in~\cref{sec:noiseless_analysis,sec:noise_analysis}, respectively.
In~\cref{sec:pedestal,sec:non_zero_n} we determine the bias in the estimation of $\mu$, its dispersion $\sigma_\mu$, and the value of $\mu$ providing the best relative precision $\sigma_\mu/\mu$ for both the pedestal events ($n=0$) and events with non-zero number of photoelectrons. 
Finally, in~\cref{sec:summary} we draw our conclusions.
\section{Introductory matter for a statistical analysis}
\label{sec:statistics}
Random variable $x$ following the probability density function $f(x)$ is denoted as $x\sim f(x)$.
If $f(x)$ depends on parameter(s) $\theta$, it is encoded as $f(x|\theta)$.

(i) The number of photons produced by a stable light source with absence of correlations and thermal noise contribution (e.g. pulsed LASER or LED) follows the binomial distribution, in general.
Given the very large number of atoms emitting the photons and small emission probability, the binomial distribution is accurately approximated by the Poisson distribution, $n_\gamma\sim P(n_\gamma|\overline{n}_\gamma)$, where  
\begin{equation}
    \label{eq:poisson}
    P(n|\mu) = \frac{\mu^n e^{-\mu}}{n!},
\end{equation}
is the Poisson probability function.

(ii) The number of photoelectrons $n$ conditional on $n_\gamma$,  $n\sim P(n|\overline{n}_\gamma\varepsilon)$.

(iii) The number $N_n$ of events  with a particular number $n$ of photoelectrons  produced in $N$  light flashes (triggers), with mean number of photons $\overline{n}_\gamma$ in each flash, hitting a photodetector with the Photon Detection Efficiency $\varepsilon$,  
$N_n\sim B(N_n,N,P(n|\overline{n}_\gamma\varepsilon))$, where 
\begin{equation}
\label{eq:binomial}
    B(k,n,p) = C^{n}_k p^k(1-p)^{n-k}, \text{ with } C^{n}_k = \frac{n!}{k!(n-k)!}
\end{equation}
is the binomial distribution.

(iv) The probability to observe $N_0$, $N_1$, $\dots$, $N_\infty$ events with zero, one,  $\dots$, $\infty$ number of photoelectrons, respectively, in $N$ light flashes is the multinomial probability function
\begin{equation}
    \label{eq:multinomial}
    M(N_0,\dots,N_\infty N, p_0,\dots, p_\infty) = \frac{N!}{N_0!\dots N_\infty!}\prod_{i=0}^\infty p_i^{N_i}
\end{equation}
for $\sum_{i=0}^\infty N_{i}=N$ and zero otherwise.
The probability $p_i$ in~\cref{eq:multinomial} is $p_i\equiv P(i|\mu)$.

A measured value $x_\text{meas}$ allows one to estimate $\theta$ and its confidence interval using the Maximum Likelihood Estimation (MLE) method.
The estimated value $\hat{\theta}$, denoted by a hat symbol above the parameter,  can be obtained by finding the maximum of the likelihood function $L$
\begin{equation}
    \label{eq:maximum_likelihood}
    \left.\frac{dL}{d\theta}\right|_{\theta=\hat{\theta}}=0, \text{ and } \left.\frac{d^2L}{d\theta^2}\right|_{\theta=\hat{\theta}}<0.
\end{equation}
An accurate determination of the confidence interval for $\theta$ generally involves Neyman's construction~\cite{Neyman:1937uhy} and an appropriate ordering principle like in the Feldman-Cousins method~\cite{Feldman:1997qc}.
We simplify the consideration by estimating the standard deviation $\hat{\sigma}_{\hat{\theta}}$ of an unbiased estimator $\hat{\theta}$ from the Fisher information:
\begin{equation}
    \label{eq:dispersion_general}
    \frac{1}{\hat{\sigma}^2_{\hat{\theta}}} = -\left.\frac{d^2\ln{L}}{d\theta^2}\right|_{\theta=\hat{\theta}}
\end{equation}
$1/\hat{\sigma}^2_{\hat{\theta}}$ in~\cref{eq:dispersion_general} is generalized to the inverse covariance matrix in case of than one parameter $\theta$.
This method fails if $d^2\ln{L}/d\theta^2=0$. 
We meet such an example in~\cref{sec:non_zero_n}.
\section{Determination of $\mu$ and $\sigma_\mu$ of a noiseless photodetector}
\label{sec:noiseless_analysis}
\subsection{Joint analysis of all peaks}
If all photoelectron peaks could be extracted, one can estimate $\mu$ and $\sigma_\mu$ from a joint analysis of all observed $N_0$, $N_1$, $\dots$, $N_\infty$ numbers of photoelectrons, in $N$ triggers.

The likelihood function $L$ appropriate for this problem reads
\begin{equation}
    \label{eq:likelihood_all_peaks}
    L(\mu) = M(N_0,\dots N_\infty, N, p_0,\dots p_\infty),
\end{equation}
where $M$ is the multinominal probability function given in~\cref{eq:multinomial}

The maximum of the likelihood in~\cref{eq:likelihood_all_peaks} occurs at 
\begin{equation}
\label{eq:mu_all_peaks}
    \hat{\mu} = \frac{1}{N}\sum_{n=0}^\infty n N_n
\end{equation}
which is an unbiased estimate of $\mu$.

Using \cref{eq:dispersion_general,eq:likelihood_all_peaks,eq:mu_all_peaks}, the relative standard deviation of these estimates reads
\begin{equation}
    \label{eq:dispersion_relative_all_peaks}
    \frac{\sigma_{\hat{\mu}}}{\hat{\mu}} = \frac{1}{\sqrt{N\hat{\mu}}}.
\end{equation}
This is the minimal relative standard deviation which could be obtained from the analysis of all peaks using the MLE.

\subsection{Analysis of the pedestal}
\label{sec:pedestal}
The mean number $\mu$ of photoelectrons can be estimated considering the pedestal events with zero number of photoelectrons~\cite{pedestal}.
This  method is often considered as a compromise between simplicity and precision of the evaluation.
Also, it is less sensitive to the model of the photodetector's response function, which could be quite complex~\cite{smirnov}, including the cross-talk in SiPM~\cite{crosstalk}.

The likelihood function $L$ for the pedestal  reads
\begin{equation}
    \label{eq:likelihood_pedestal}
    L(\mu) = B(N_0,N,P(0|\mu)),
\end{equation}
where $B$ and $P$ are the Binomial and Poisson probability functions given by~\cref{eq:binomial,eq:poisson}, respectively.

The solution of~\cref{eq:maximum_likelihood} for $n=0$ and $p_0\equiv P(0|\mu)=e^{-\mu}$ (the corresponding estimate is denoted as $\hat{p}_0$ in what follows) reads
\begin{equation}
\label{eq:mu_0_estimate}
    \hat{\mu}_0 = -\ln{\hat{p}_0} = -\ln{\frac{N_0}{N}},
\end{equation}
where the subscript $0$ in $\hat{\mu}_0$ indicates the method used to estimate $\mu$ ($n=0$, or zero photoelectrons).

The case of $N_0=0$ does not correspond to the maximum of $L$ in~\cref{eq:likelihood_pedestal}.
Therefore, no estimation of $\hat{\mu}_0$ is possible if $N_0=0$. 
In this case one would be able to determine the confidence interval for $\mu$ setting up an appropriate confidence level $\alpha$
\begin{equation}
 \label{eq:N_0_0_CL}
\left(-\ln\left[1-(1-\alpha)^{1/N}\right],+\infty\right).
\end{equation}

One can see that 
\begin{equation}
    \label{eq:mu_0_estimate_limit}
    \lim_{N\to\infty} \hat{\mu}_0 = \mu.
\end{equation}
Let us prove that the estimate in~\cref{eq:mu_0_estimate} is biased for a fixed $N$.
The mean value of $\hat{\mu}_0$  obtained as an average over $M$ experiments in the limit $M\to\infty$
\begin{equation}
    \label{eq:mu0_mean_1}
    E\left[\hat{\mu}_0\right] = \lim_{M\to\infty}\frac{1}{M}\sum_{k=1}^M \hat{\mu}_{0k} = - \lim_{M\to\infty}\frac{1}{M}\sum_{k=1}^M \ln{\frac{N_{0k}}{N}}
\end{equation}
differs from the estimate $\hat{\mu}_0$ based on the joint analysis of all $M$ experiments.

For the latter, one should generalize the likelihood in~\cref{eq:likelihood_pedestal} appropriately
\begin{equation}
    \label{eq:likelihood_joint}
    L(\mu) = \lim_{M\to\infty}\prod_{k=1}^M B(N_{0k},N,P(0|\mu))
\end{equation}
for which the solution of~\cref{eq:maximum_likelihood} reads
\begin{equation}
\label{eq:mu_0_estimate_2}
    \hat{\mu}_0 = -\lim_{M\to\infty}\ln{\left(\frac{1}{M}\sum_{k=1}^M\frac{N_{0k}}{N}\right)} = \mu=\hat{\mu}_0(\overline{P}),
\end{equation}
where the last equality is according to~\cref{eq:mu_0_estimate_limit}.
$E\left[\hat{\mu}_0\right]$ from~\cref{eq:mu0_mean_1} differs from $\hat{\mu}_0$ in~\cref{eq:mu_0_estimate_2} unless $N\to\infty$.

According to the Jensen's inequality~\cite{jensen1906} for a convex function for a finite~$M$
\begin{equation}
    \ln{\left(\frac{1}{M}\sum_{k=1}^M\frac{N_{0k}}{N}\right)} \ge \frac{1}{M}\sum_{k=1}^M \ln{\frac{N_{0k}}{N}}.
\end{equation}
Therefore, 
\begin{equation}
    \label{eq:bias_jensen}
    E\left[\hat{\mu}_0\right] \ge \mu
\end{equation}
and the estimate in~\cref{eq:mu_0_estimate} is biased.

Let us estimate the bias
\begin{equation}
    \label{eq:bias1}
    \beta(\hat{\mu}_0)  = E\left[\hat{\mu}_0\right] - \mu
\end{equation}
expanding $\hat{\mu}_0$ to the second order in $\hat{p}_0$ around its mean value $\overline{p}$
\begin{equation}
    \label{eq:bias2}
    \begin{aligned}
    \beta(\hat{\mu}_0)  & \approx E\left[\hat{\mu}_0(\overline{p}_0)-\mu +\left.\frac{d\hat{\mu}_0}{d\hat{p}_0}\right|_{\hat{p}_0=\overline{p}_0}(\hat{p}_0-\overline{p}_0)
    +\frac{1}{2}\left.\frac{d^2\hat{\mu}_0}{d\hat{p}^2_0}\right|_{\hat{p}_0=\overline{p}_0}(\hat{p}_0-\overline{p}_0)^2\right] \\
    & = \frac{1}{2}\left.\frac{d^2\hat{\mu}_0}{d\hat{p}^2_0}\right|_{\hat{p}_0=\overline{p}_0}\sigma^2_{\hat{p}_0} 
    \approx \frac{1-\hat{p}_0}{2N\hat{p}_0} =  \frac{e^{\hat{\mu}}-1}{2N}\ge 0.
    \end{aligned}
\end{equation}

The bias in~\cref{eq:bias2} $\beta(\hat{\mu}_0)\ge 0$ is in agreement with~\cref{eq:bias_jensen} and it vanishes for $N\to\infty$.

The dispersion of the bias-corrected estimation $\mu_0= \hat{\mu}_0-\beta(\hat{\mu}_0)$ can be obtained at point $\mu = \mu_0$ with the help of~\cref{eq:dispersion_general,eq:likelihood_pedestal}
\begin{equation}
    \label{eq:disper_mu0}
    \hat{\sigma}_{\mu_0}^2 = \frac{(1-e^{-\mu_0})^2}{e^{-\mu_0}(N-N_0)} = \frac{(e^{\mu_0}-1)}{N}S_0^2,
\end{equation}
where 
\begin{equation}
    \label{eq:dispersion_factor_mu0}
    S_0^2 = \frac{(e^{\mu_0}-1)}{e^{\mu_0}-e^{-\beta(\hat{\mu}_0)}} \approx 1-\frac{1}{2N}. 
\end{equation}
The approximate equality in~\cref{eq:dispersion_factor_mu0} corresponds to $N\gg 1$. 
Finally, the relative standard deviation of an unbiased estimate
\begin{equation}
    \label{eq:dispersion_mu0}
    \frac{\hat{\sigma}_{\mu_0}}{\mu_0} =  \frac{1}{\sqrt{N\mu_0}}  \sqrt{\frac{e^{\mu_0}-1}{\mu_0}}S_0
\end{equation}
is a product of three factors.

The first factor is equal to  the minimum possible relative standard deviation of $\hat{\mu}$ obtained from the joint analysis of all peaks, as can be seen in~\cref{eq:dispersion_relative_all_peaks}.
The second factor,  always larger than the first, reflects the fact that only partial information about the number of photoelectrons is used in the pedestal method. 
At $\mu\to~0$ the first factor approaches unity and at $\mu\gg~1$ it grows exponentially, manifesting that the pedestal is far from  optimal in that limit.
The third factor is a correction of the order of one.

\begin{figure}[!ht]
   \centering
  \includegraphics[width=230px]{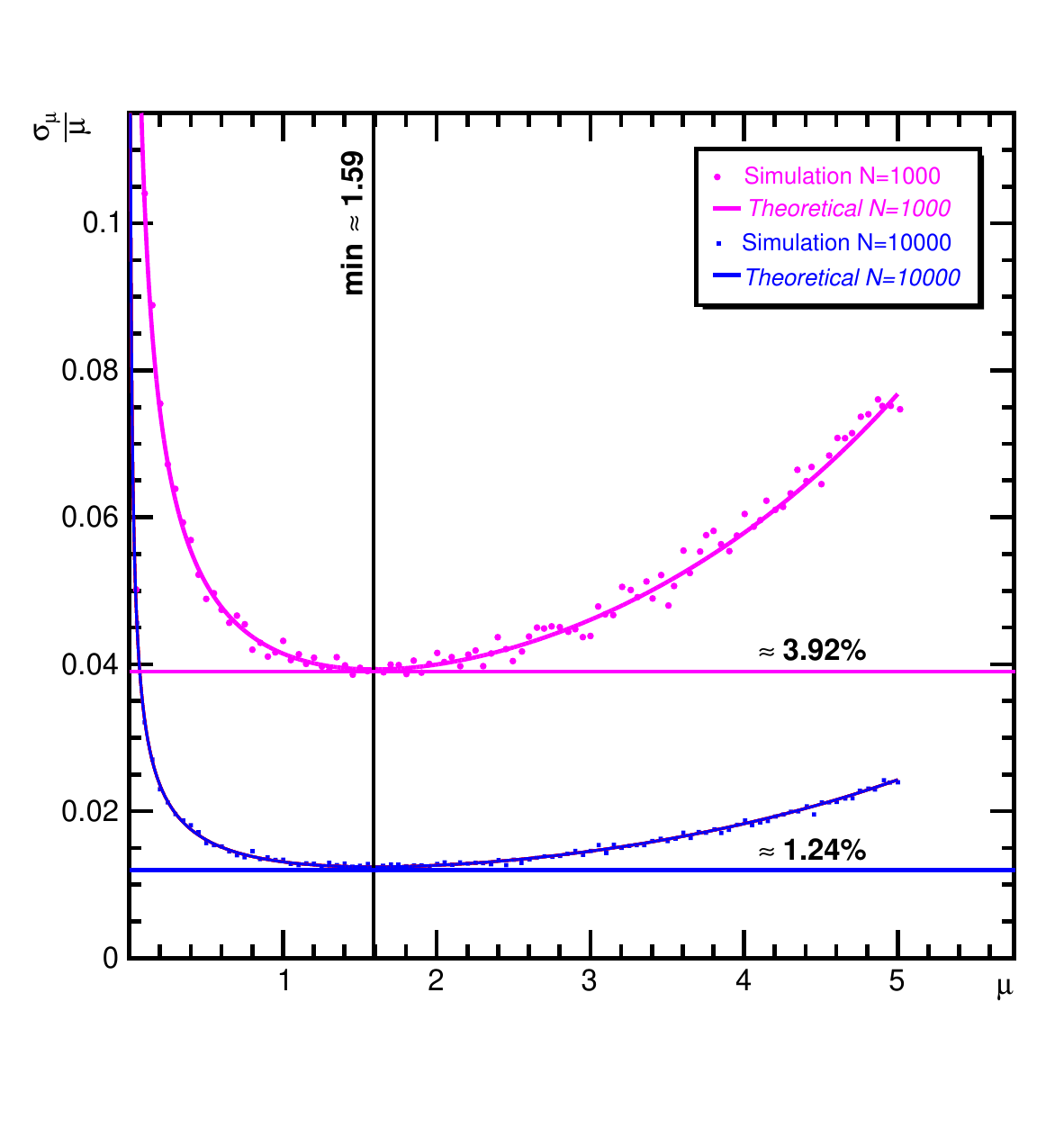} \\ 
   \caption{Relative dispersion $\sigma_{\mu_0}/\mu_0$ of $\mu$ estimation by the pedestal method  as function of  $\mu$ for $N=1000$ (magenta) and $N=10000$ (blue) triggers. The points correspond to synthetic experiments with $M=1000$. The curves correspond to~\cref{eq:dispersion_mu0}  in which $\mu_0\to \mu$ and $S_0\to 1$.}
  \label{fig:fig1}
\end{figure}

The product of these factors has a minimum at $\mu\simeq 1.59$, as can be seen in~\cref{fig:fig1}.
This is the optimal value of $\mu$ which provides the best estimation of $\mu$ around this value using the pedestal method with fixed statistics.
The relative standard deviation of $\mu$ at $\mu\simeq 1.59$ ($p_0\simeq 0.204$) is about $1.57$ times larger than that obtained from a joint analysis of all possible peaks.

To verify our calculations, simulations of synthetic experiments were performed with the following algorithm.

(i) For every experiment $j=(1,M)$ generate $N$ numbers $n_i\sim P(n_i|\mu)$.  

(ii) Count $N_{0}$ -- the number of cases where $n_i=0$.
If $N_{0}\ne 0$, estimate $\hat{\mu}_{0j}$ with help of~\cref{eq:mu_0_estimate}. 
Otherwise, skip this experiment since no estimate is possible.
This approach would be practical for $\mu\le 5$ and $N\ge 1000$ since the probability to observe $N_0=0$ is $(1-e^{-\mu})^{N}\approx 10^{-3}$ for $\mu=5$ and $N=10^3$, which corresponds to about 1 experiment out of $M=1000$ where an estimation is not possible. 
(iii) Estimate the bias $\beta(\hat{\mu}_{0j})$ using~\cref{eq:bias2} and evaluate the unbiased estimation $\mu_{0j}= \hat{\mu}_{0j}-\beta(\hat{\mu}_{0j})$. 

Calculate the mean and its variance of $\mu_0$ distribution for every $\mu \in (0.05,5)$ with step 0.05.

Compare the relative standard deviation of $\mu$ to~\cref{eq:dispersion_mu0} in which $\mu_0\to \mu$ and $S_0\to 1$ as displayed in~\cref{fig:fig1}. 
\subsection{Generalization for $n>0$}
\label{sec:non_zero_n}
The likelihood function corresponding to the number of photoelectrons $n\ne~0$ reads
\begin{equation}
    \label{eq:likelihood_n_peak}
    L(\mu) = B(N_n,N,P(n|\mu)).
\end{equation}
The cases of $N_n=0$, $\mu=n$ ($\hat{p}_n < p_n^\text{crit} = P(n|n)$) do not correspond to the maximum of $L$ in~\cref{eq:likelihood_n_peak}, otherwise the estimator formally reads
\begin{equation}
\label{eq:mu_n_estimate}
\hat{\mu}_{n} = - nW(-\frac{\sqrt[n]{\hat{p}_{n}n!}}{n}), 
\end{equation}
where $\hat{p}_n\equiv P(n|\hat{\mu})$, $W$ is the  Lambert $W$ function \cite{Lambert} and $\hat{\mu}_{n} \ne n$. If $\hat{p}_n \ge p_n^\text{crit}$ maximum of $L$ in \cref{eq:likelihood_n_peak} is obtained at $\hat{\mu}_n=n$.

The Lambert $W$ function is double-valued, thus yielding an ambiguity in $\mu$ determination.
Additional inputs are required in order to resolve it.
In practice, one could perform complementary measurements with a different light intensity and observe the change of the estimator $\hat{\mu}_n$.
A fake estimator  could be detected if $\hat{\mu}_n$ decreases (increases) when the light intensity increases (decreases)
\footnote{Alternatively, one could evaluate $\hat{\mu}$ by using other photoelectron peaks or to use other properties of the Poisson distribution (e.g.~relation between variance and mean which is in ideal case expressed by~\cref{eq:dispersion_relative_all_peaks}).}.

The bias of $\hat{\mu}_n$  can be expressed  similarly to~\cref{eq:bias2}
\begin{equation}
    \label{eq:bias2_mu_n}
    \beta(\hat{\mu}_n) \approx \frac{p^{-1}_n-1}{2N}\times\frac{(n-(n-\hat{\mu}_n)^2)\hat{\mu}_n}{(n-\hat{\mu}_n)^3}.
\end{equation}
The relative standard deviation of the bias-corrected estimator $\mu_n= \hat{\mu}_n-\beta(\hat{\mu}_n)$ can be obtained with help of~\cref{eq:dispersion_general,eq:likelihood_n_peak}
\begin{equation}
\label{eq:dispersion_mu_n}
\frac{\hat{\sigma}_{\mu_n}}{\mu_n} = S_n \frac{1}{\sqrt{N}}\sqrt{\frac{{p}^{-1}_n-1}{(\mu_n-n)^{2}}}
\end{equation}
where
\begin{equation}
 \label{eq:dispersion_factor_mu_n}
S_n^2 \approx 1 - \frac{1}{2N} + \frac{1}{2N}\left[\frac{n(\mu_n-n)^2p^{-1}_n+n^2(p^{-1}_n-1)}{(\mu_n-n)^4}\right]
\end{equation}
is an approximation for small $\beta(\hat{\mu}_n)$ and $p_n\equiv P(n|\mu_n)$ in~\cref{eq:bias2_mu_n,eq:dispersion_mu_n}. 
One can see that \cref{eq:bias2,eq:dispersion_mu0} are a special case of~\cref{eq:bias2_mu_n,eq:dispersion_mu_n} with $n=0$.

\begin{figure}[ht!]
   \centering
   \includegraphics[width=230px]{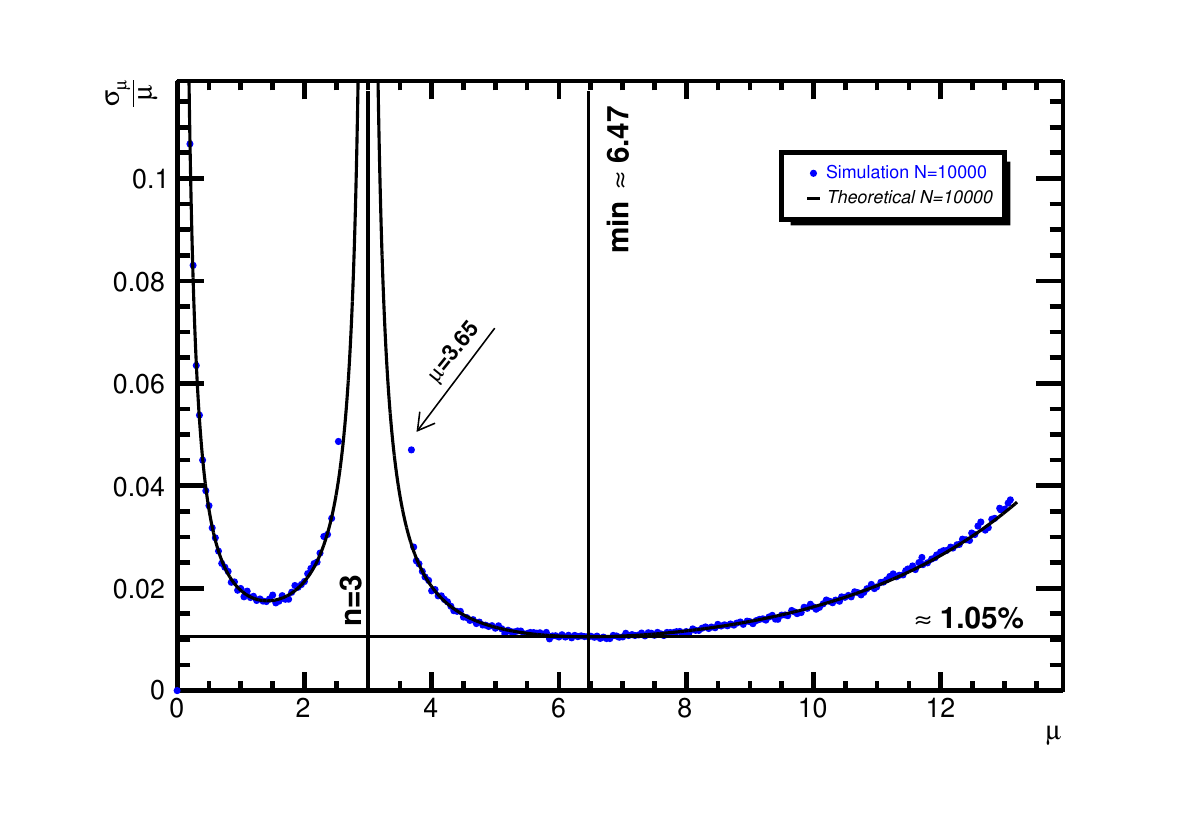} \\ 
   \caption{Relative dispersion $\sigma_{\mu_3}/\mu_3$ of $\mu$ estimation by the $3^{rd}$ peak  as function of  $\mu$ for  $N~=10000$  triggers. 
   The points correspond to synthetic experiments with $M=1000$. 
   The curve corresponds to~\cref{eq:dispersion_mu_n}  in which $n=3$, $\mu_n\to \mu$ and $S_n$ is given by~\cref{eq:dispersion_factor_mu_n}.} 
  \label{fig:fig2}
\end{figure}

As an illustration, we display in~\cref{fig:fig2} the expected relative dispersion $\sigma_{\mu_3}/\mu_3$ of $\mu$ estimation by the $3^{rd}$ peak  as function of  $\mu$ for  $N~=10000$  triggers. 
One can observe two minima in the  relative standard deviation curve.
The right minimum is deeper and occurs at $\mu_\text{opt}\simeq 6.47$ ($p_3 \simeq 0.070$) where the relative uncertainty is expected to be about $1.05\%$ for $N~=10000$  triggers.
Also, one might observe a discontinuity at $\mu \to n=3$.
At this point the second derivative of $\ln{L}$ vanishes and the estimate of variance of $\mu$ with help of~\cref{eq:dispersion_general} becomes impossible.
Around this point the bias $\beta(\hat{\mu}_n)$ is also significant.
Therefore~\cref{eq:bias2_mu_n,eq:dispersion_factor_mu_n,eq:dispersion_mu_n} are incorrect. 
A demonstration of wrong estimation is shown by a point $\mu=3.65$ in~\cref{fig:fig2}.
One has to reconsider the determination of the confidence interval for $\mu$ using the full Neyman's construction, which is beyond of scope of the current manuscript.
As one can observe, working with $n\ne 0$ is significantly more complicated with respect to $n=0$ case.

\section{Determination of $\mu$ and $\sigma_\mu$ with a noisy photodetector}
\label{sec:noise_analysis}
Let us consider a simplified noise model, assuming the noise pulses to be uncorrelated.
The number $m$ of dark pulses in the time window $\tau$ could be approximated with a Poisson distribution $m\sim P(m|\lambda)$, where $\lambda=\tau R$ and $R$ is an average dark count rate, reciprocal to seconds.
The number $n$ in the resulting spectrum (signal+noise) due to both signal and noise again follows the Poisson distribution $n\sim P(n|\xi)$, where $\xi = \mu + \lambda$.

To proceed further, $\lambda$ must be estimated.
In practice, this can be done by measuring the dark noise when switching off the light generator.
The estimator $\hat{\lambda}_m$ corresponds to the $m^\text{th}$ peak. 
Estimators $\hat{\xi}_n$ for the resulting spectrum  and $\hat{\lambda}_m$ for the dark pulses spectrum (noise) could be found to be similar to $\hat{\mu}_n$ (see \cref{eq:mu_0_estimate,eq:mu_n_estimate}).
The likelihood function of combined measurement reads
\begin{equation}
    \label{eq:likelihood_noise}
    L(\xi,\lambda)=B(N_n,N,P(n|\xi))\times B(D_m,D,P(m|\lambda)) = L_1(\mu,\lambda)\times L_2(\lambda)
\end{equation}
where $N$,$N_n$ gives total and  $n^\text{th}$ peak's number of events in the resulting spectrum and $D$,$D_m$ the total and $m^\text{th}$ peak's number of events in the dark spectrum. 

In the simplest case of $n=m=0$ the maximum of $L$ can be found at
\begin{equation}
    \label{eq:mu_00}
   \hat{\mu}_{0,0} = - \ln{\left(\frac{N_0}{N}\cdot\frac{D}{D_0}\right)} = -\ln{\left(\frac{\hat{p}_{\xi_0}}{\hat{p}_{\lambda_0}}\right)} = \hat{\xi}_0 - \hat{\lambda}_0
\end{equation}
accounting that $p_{\xi_0} \equiv P(0|\xi)$ and $p_{\lambda_0} \equiv P(0|\lambda)$. 
Index $(0,0)$ refers to an estimation by pedestals both in the dark and resulting spectra. 
Solutions $N_0=0$ and $D_0 = 0$ do not correspond to a maximum of $L$ in \cref{eq:likelihood_noise}.
To evaluate an unbiased estimator for $\mu_{0,0}$ we may express the bias as
\begin{equation}
    \label{eq:bias_mu_00}
   \beta(\hat{\mu}_{0,0}) = E[\hat{\mu}_{0,0}] - \mu =  E[\hat{\xi}_0 - \hat{\lambda}_0] - (\xi - \lambda)
   = \beta(\hat{\xi}_0) - \beta(\hat{\lambda}_0).
\end{equation}
Then, $\mu_{0,0}=\xi_0-\lambda_0$. 
The calculation of biased corrected estimators $\xi_0$ and $\lambda_0$ is done in a similar fashion to $\mu_0$ (see \cref{eq:bias1}). 

Since, the likelihood in~\cref{eq:likelihood_noise} depends on more than one parameter, ~\cref{eq:dispersion_general} must be generalized to the covariance matrix for estimates of parameters $\mu,\lambda$.
Let us denote, for the sake of compactness, $\hat{\sigma}_{\xi_0}^2$ and $\hat{\sigma}_{\lambda_0}^2$  as the variances of unbiased estimators $\xi_0$ and $\lambda_0$ in signal+noise and in noise spectra, respectively.
The inverse covariance matrix of estimated parameters $\mu,\lambda$ reads
\begin{equation}
\label{eq:inverse_cov}
  V^{-1} = -
\begin{pmatrix}
\frac{\partial^2\ln{L}}{\partial \mu^2} & \frac{\partial^2\ln{L}}{\partial \mu \partial \lambda}\\
\frac{\partial^2\ln{L}}{\partial \lambda \partial \mu} & \frac{\partial^2\ln{L}}{\partial \lambda^2}\\
\end{pmatrix} =
\begin{pmatrix}
\hat{\sigma}_{\xi_0}^{-2} & \hat{\sigma}_{\xi_0}^{-2} \\
\hat{\sigma}_{\xi_0}^{-2}  & \hat{\sigma}_{\xi_0}^{-2} + \hat{\sigma}_{\lambda_0}^{-2}  \\ 
\end{pmatrix}.
\end{equation}
Inverting~\cref{eq:inverse_cov} one gets
\begin{equation}
\label{eq:cov}
  V = 
\begin{pmatrix}
\text{Var}(\mu) & \text{Cov}(\mu, \lambda)\\
\text{Cov}( \lambda, \mu) & \text{Var}(\lambda) \\
\end{pmatrix} = \begin{pmatrix}
\hat{\sigma}_{\xi_0}^{2} + \hat{\sigma}_{\lambda_0}^{2} &-\hat{\sigma}_{\xi_0}^{2} \\ -\hat{\sigma}_{\xi_0}^{2} &
\hat{\sigma}_{\lambda_0}^{2}\\  
\end{pmatrix}
\end{equation}
The variance of $\hat{\mu}_{0,0}$ in the signal+noise spectrum can be read from~\cref{eq:cov} 
\begin{equation}
    \label{eq:var_mu_signal_noise}
     \text{Var}(\mu) \equiv \hat{\sigma}_{\mu_{0,0}}^2  = \hat{\sigma}_{\xi_0}^2 + \hat{\sigma}_{\lambda_0}^2.
\end{equation}
It is instructive to evaluate $\text{Var}(\mu)$ by a different method. 
One can make it without referring to the covariance matrix  in~\cref{eq:cov}, \textit{profiling} $\ln{L}$ over $\lambda$.
The profiling consists of the following steps.

(i) For any $\mu$ find $\hat{\hat{\lambda}}(\mu)$, which is a function of $\mu$, such that
\begin{equation}
    \label{eq:likelihood_profiling_1}
   \left. \frac{\partial }{\partial\lambda}\ln{L}(\mu,\lambda)\right|_{\lambda=\hat{\hat{\lambda}}(\mu)}= 0.
\end{equation}

(ii) Replace $\lambda$ by $\hat{\hat{\lambda}}(\mu)$ in $\ln{L}(\mu,\lambda)$ and find $\hat{\mu}$ as solution of
\begin{equation}
    \label{eq:likelihood_profiling_2}
    \left.\frac{d}{d\mu}\ln{L}(\mu,\hat{\hat{\lambda}}(\mu))\right|_{\mu=\hat{\mu}}= 0.
\end{equation}

(iii) Find $\text{Var}(\mu)$ using~\cref{eq:dispersion_general,eq:likelihood_profiling_2}.

Performing steps (i)-(iii) one obtains exactly the same $\text{Var}(\mu)$ as in~\cref{eq:var_mu_signal_noise}.

Approximating  the dispersion factor $S_0\approx1$ (see~\cref{eq:dispersion_factor_mu0}) and assuming $N=D$  the relative standard deviation of  $\hat{\mu}_{0,0}$  reads
\begin{equation}
    \label{eq:dispersion_mu_00_noise}
      \frac{\hat{\sigma}_{\mu_{0,0}}}{\mu_{0,0}} \approx \frac{1}{\sqrt{N}}\sqrt{\frac{e^{\xi_0}+ e^{\lambda_0}-2}{(\xi_0-\lambda_0)^2}}.
\end{equation}

We cross-check~\cref{eq:dispersion_mu_00_noise} by a simulation.

(i) For every experiment $j=(1,M)$ simultaneously generate $N$ numbers $l\sim P(l|\mu)$ and  $k\sim P(k|\lambda),$
calculate $n=l+k$. 
$n\sim P(n|\xi)$, where $\xi=\mu+\lambda$.

(ii) For every experiment generate $N$ numbers $m\sim P(m|\lambda)$ to synthesize the noise spectrum~\footnote{Two noise generators $P(k|\lambda)$ and $P(m|\lambda)$ are different sequences and independent.} 

(iii) Count $N_{0}$, which gives the number of cases where $n=0$ in signal+noise spectrum and $D_0$,  where $m=0$ in the noise spectrum.
If $N_{0}, D_0 \ne 0$, estimate $\hat{\xi}_{0j}$ and $\hat{\lambda}_{0j}$  with help of~\cref{eq:mu_0_estimate}. 
Otherwise, skip this experiment since no estimate is possible.

(iv) Estimate the biases $\beta(\hat{\xi}_{0j})$, $\beta(\hat{\lambda}_{0j})$ using~\cref{eq:bias2} and evaluate the unbiased estimation $\mu_{0,0j}= \xi_{0j} -  \lambda_{0j} = (\hat{\xi}_{0j} - \beta(\hat{\xi}_{0j})) - (\hat{\lambda}_{0j} - \beta(\hat{\lambda}_{0j}))$. 

Calculate the mean $E[\mu_{0,0}]$ and its standard deviation $\hat{\sigma}_{\mu_{0,0}}$ as square root of variance   for every $\mu \in (0.05,5)$ with step size $0.05$.

\begin{figure}[!ht]
   \centering
  \includegraphics[width=230px]{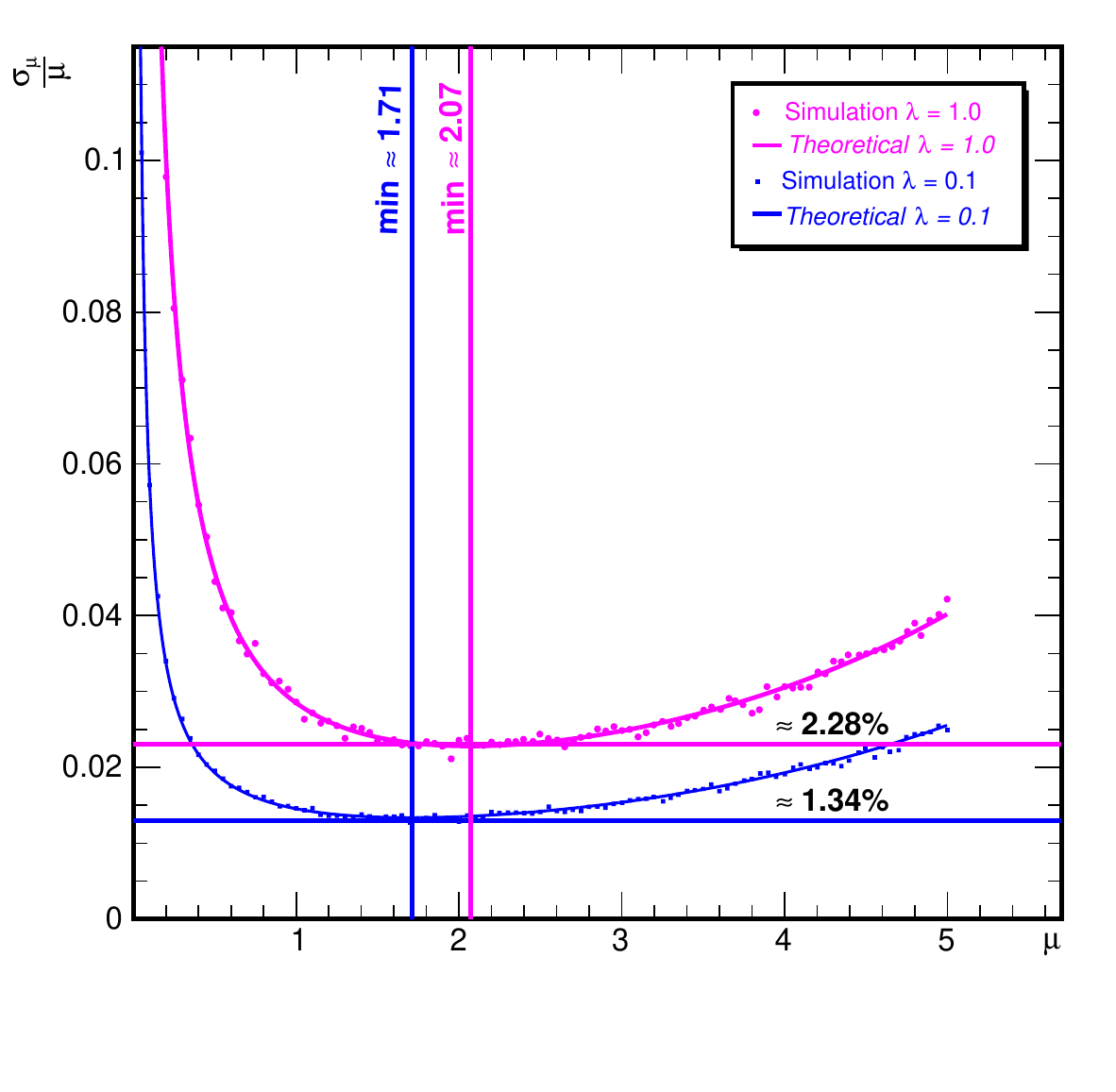} \\ 
   \caption{Relative dispersion $\sigma_{\mu_{0,0}}/\mu_{0,0}$ of $\mu$ estimation by the pedestal method for signal+noise and noise spectra  as function of  $\mu$ for $N=10000$ triggers for two different noise levels $\lambda=1.0$ (magenta) and $\lambda=0.1$ (blue). 
   The points correspond to synthetic experiments with $M=1000$. 
   The curves correspond to~\cref{eq:dispersion_mu00_theor}.}
  \label{fig:fig3}
\end{figure}

The relative standard deviation of $\mu$ to~\cref{eq:dispersion_mu_00_noise} in which $\xi_0\to \xi=\mu+\lambda$ and $\lambda_0\to \lambda$ 
\begin{equation}
    \label{eq:dispersion_mu00_theor}
 \frac{\hat{\sigma}_{\mu_{0,0}}}{E[\mu_{0,0}]} \approx \frac{1}{\sqrt{N}}\sqrt{\frac{e^{\mu+\lambda}+ e^{\lambda}-2}{\mu^2}}
 \end{equation}
is displayed in~\cref{fig:fig3}.
One can see that the resolution degrades with increasing noise ($\lambda$) and the optimum $\mu$ shifts to higher values with respect to a noiseless photodetector. 
We calculate the parameters for some realistic case and list them in~\cref{tab:tab1}. 
To reach the same statistical precision as in the noiseless case ($\lambda=0$) the acquisition time must be increased by factor $F^2$ at the optimal point $\mu_\text{opt}$.
\begin{table}
\centering
\begin{tabular}{lllll}
$\tau=100$ ns, $N=10000$, $n=0$, $m=0$              &                             &                              &       & \multicolumn{1}{l} {}              \\ \hline
\multicolumn{1}{|l|}{R, $\text{sec}^{-1}$} & \multicolumn{1}{l|}{$\lambda$} & \multicolumn{1}{l|}{$\mu_\text{opt}$} & \multicolumn{1}{l|}{$\sigma_{\mu}/\mu$ ,\%} & \multicolumn{1}{l|}{$F$}\\ \hline
\multicolumn{1}{|l|}{0}          & \multicolumn{1}{l|}{0}       & \multicolumn{1}{l|}{1.59}        & \multicolumn{1}{l|}{1.24}    & \multicolumn{1}{l|}{1.00}        \\ \hline
\multicolumn{1}{|l|}{$10^{3}$ (1 k$\text{sec}^{-1}$)}          & \multicolumn{1}{l|}{$10^{-4}$}       & \multicolumn{1}{l|}{1.59}        & \multicolumn{1}{l|}{1.24}  & \multicolumn{1}{l|}{1.00}          \\ \hline
\multicolumn{1}{|l|}{$10^4$ (10 k$\text{sec}^{-1}$)}          & \multicolumn{1}{l|}{$10^{-3}$}       & \multicolumn{1}{l|}{1.60}        & \multicolumn{1}{l|}{1.24}   & \multicolumn{1}{l|}{1.00}         \\ \hline
\multicolumn{1}{|l|}{$10^5$ (100 k$\text{sec}^{-1}$)}          & \multicolumn{1}{l|}{$10^{-2}$}       & \multicolumn{1}{l|}{1.61}        & \multicolumn{1}{l|}{1.25}  & \multicolumn{1}{l|}{1.01}          \\ \hline
\multicolumn{1}{|l|}{$10^6$ (1 M$\text{sec}^{-1}$)}          & \multicolumn{1}{l|}{$10^{-1}$}       & \multicolumn{1}{l|}{1.71}        & \multicolumn{1}{l|}{1.34}  & \multicolumn{1}{l|}{1.08}          \\ \hline
\multicolumn{1}{|l|}{$10^7$ (10 M$\text{sec}^{-1}$)}          & \multicolumn{1}{l|}{1}       & \multicolumn{1}{l|}{2.07}        & \multicolumn{1}{l|}{2.28}   & \multicolumn{1}{l|}{1.83}         \\ \hline
\end{tabular}
\caption{Noise case: dark rate $R$, average number of dark pulses per trigger $\lambda$, optimal light intensity $\mu_\text{opt}$, best relative accuracy $\sigma_{\mu}/\mu$ and noise factor $F$, which is defined by a ratio of the best accuracy with noise to the noiseless $(\lambda=0)$.}
\label{tab:tab1}
\end{table}
\section{Summary}
Commissioning of a large number of photodetectors requires  optimization of the time needed to characterize a single unit.
Depending on a chosen characterization method, the minimum time can be achieved selecting an optimal intensity of light illuminating the photodetector.
In this work, we propose an optimal light intensity and estimate a statistical accuracy in determination of the photon detection efficiency PDE, which can be achieved by a particular method, based on measuring $n^{\text{th}}$ photoelectron peak.

As a practical illustration of our strategy, let us consider a PMT scanning station used for the characterization of large PMTs~\cite{JUNO-anfimov}.
The average number of photoelectrons $\mu$ is estimated with help of the pedestal events, which are events with zero number of photoelectrons as a response to the light illumination.
The DAQ of the station is provided by the DRS4 evaluation board, which can afford about 500 events/sec at most.
In a regime of a detailed PMT characterization, there are 168 points of light incidence over the PMT surface.
Each point accumulates about 10 thousands of events, pushing the total time needed to scan the entire PMT to about one hour.
The best statistical accuracy of about 1.2\% in $\mu$ estimation with $N=10^4$ events can be achieved at  $\mu_\text{opt}=1.6$ photoelectrons if noise contribution is negligible, as can be seen from \cref{fig:fig1}.

There are at least $1.54\cdot 10^{4}$  events  required for pedestal method to improve the statistical accuracy to $1\%$. 
In the presence of noise the optimal light intensity $\mu_\text{opt}$ shifts towards a higher value, while accuracy  in $\mu$  determination degrades by the noise factor $F$, as can be seen in~\cref{tab:tab1}.
For large enough $\lambda$ the statistics of trigger events should be increased by a factor $F^2$ in order to reach the same accuracy  as in a case of noiseless PMT.
In practice, $\lambda>0.1$ is a minimum value requiring an increase of the number of trigger events.

For $R<$100 $\text{ksec}^{-1}$ and trigger window $\tau<$ 100 ns, $\lambda<10^{-2}$ which has a negligible impact for $\mu$ determination as can be seen from~\cref{tab:tab1}.
One can see from fig.1 that functions are approximately flat withing a range from 1~to~2 photoelectrons.
To obtain the best precision we propose to adjust the light intensity for tested photosensors within this range.
In general, a method based on a single $n^{\text{th}}$ photoelectron peak evaluation, estimates $\mu$ with a bias which should be corrected.

\begin{figure}[!ht]
   \centering
   \includegraphics[width=230px]{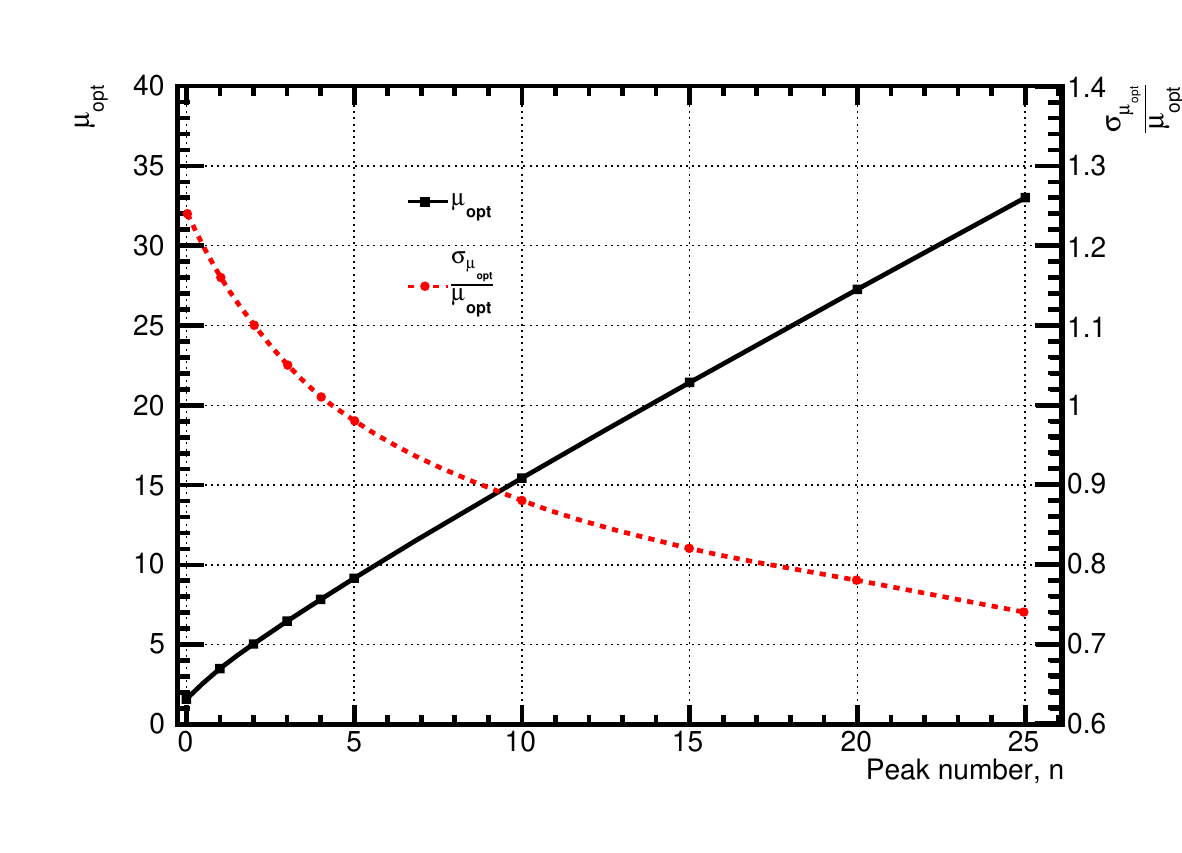} \\ 
   \caption{Optimal light intensity $\mu_\text{opt}$ and best accuracy at optimum $\sigma_{\mu_\text{opt}}/\mu_\text{opt}$ as functions of $n$, used in $\mu$ determination.} 
  \label{fig:fig4}
\end{figure}

For a visual clarity and as a short summary,~\cref{fig:fig4} displays an optimal light intensity $\mu_\text{opt}$ and expected accuracy  $\sigma_{\mu_\text{opt}}/\mu_\text{opt}$ as functions of $n$, used in $\mu$ determination.
\label{sec:summary}
\section{Acknowledgments}
The authors are indebted to Prof.~Dr.~D.~V.~Naumov  for his great help in the preparation of this paper. 
We are also thankful to O.~Smirnov, C.~Kullenberg and S.~Gursky for reading the manuscript and making a number of useful comments. This research did not receive any specific grant from funding agencies in the public, commercial, or
not-for-profit sectors.
\label{sec:acknowledgments}
\bibliography{manuscript}
\end{document}